\documentclass[journal]{IEEEtran}
\IEEEoverridecommandlockouts
\usepackage{booktabs}
\usepackage{siunitx}
\usepackage{cite}
\usepackage{amsmath,amssymb,amsfonts}
\usepackage{graphics, framed}
\usepackage{graphicx}
\usepackage{epsfig}
\usepackage{epstopdf}
\usepackage{url}
\usepackage{multimedia}
\usepackage{paralist}
\usepackage[printonlyused]{acronym}
\usepackage{units}
\usepackage{algorithmic}
\usepackage{textcomp}
\usepackage{float}
\usepackage{tabularx}
\usepackage{balance}
\usepackage{subfigure}
\usepackage{xcolor}
\def\BibTeX{{\rm B\kern-.05em{\sc i\kern-.025em b}\kern-.08em
T\kern-.1667em\lower.7ex\hbox{E}\kern-.125emX}}
\usepackage{mathtools}
\usepackage{cuted} 
\usepackage[export]{adjustbox}

\usepackage{dblfloatfix}
\usepackage{lipsum}
\usepackage{color,soul}
\usepackage{multirow}
\usepackage[overload]{textcase}
\newcommand{\FGR}[1]{Fig.~\ref{#1}}

\newcommand{\SEC}[1]{Section~\ref{#1}}

\acrodef{5G}[5G]{5\textsuperscript{th}-Generation}
\acrodef{BW}[BW]{bandwidth}
\acrodef{CW}[CW]{continuous wave}
\acrodef{D2D}[D2D]{device-to-device}
\acrodef{dB}[dB]{decibel}
\acrodef{dBi}[dBi]{decibel isotropic}
\acrodef{dBm}[dBm]{decibel over a milliwatt}
\acrodef{Gbps}[Gbps]{gigabit per second}
\acrodef{GHz}[GHz]{gigahertz}
\acrodef{THz}[THz]{Terahertz}
\acrodef{RIS}[RIS]{reconfigurable intelligent surface}
\acrodef{GM}[GM]{Gamma mixture}
\acrodef{PSK}[PSK]{phase shift keying}
\acrodef{QAM}[QAM]{quadrature amplitude modulation}
\acrodef{AWGN}[AWGN]{additive white Gaussian noise}
\acrodef{SNR}[SNR]{signal-to-noise ratio}
\acrodef{AF}[AF]{amplitude-and-forward}
\acrodef{MIMO}[MIMO]{multiple-input multiple-output}
\acrodef{mMIMO}[mMIMO]{massive-multiple-input multiple-output}
\acrodef{SDN}[SDN]{Software-defined network}
\acrodef{SON}[SON]{self-organizing network}
\acrodef{hetnet}[HetNet]{heterogeneous network}
\acrodef{FSO}[FSO]{free-space optics}
\acrodef{UM-MIMO}[UM-MIMO]{ultra-massive-MIMO}
\acrodef{AP}[AP]{access point}
\acrodef{UE}[UE]{user equipment}
\acrodef{NFP}[NFP]{networked flying platform}
\acrodef{UAV}[UAV]{unmanned aerial vehicle}
\acrodef{HAP}[HAPS]{high-altitude platform station}
\acrodef{LEO}[LEO]{low-earth orbit}
\acrodef{BAN}[BAN]{body area network}
\acrodef{WLAN}[WLAN]{wireless local area network}
\acrodef{QoS}[QoS]{quality of service}
\acrodef{TCS}[TCS]{thermal control system}
\acrodef{QCL}[QCL]{quantum cascade laser}
\acrodef{CMOS}[CMOS]{complementary metal-oxide semiconductor}
\acrodef{V-HetNet}[V-HetNet]{vertical heterogeneous network}
\acrodef{DL}[DL]{Deep learning}
\acrodef{DRL}[DRL]{deep reinforcement learning}
\acrodef{FDTD}[FDTD]{Finite-difference time-domain}
\acrodef{FEM}[FEM]{finite element method}
\acrodef{MoM}[MoM]{method of moments}
\acrodef{VNA}[VNA]{vector network analyzer}
\acrodef{CS}[CS]{channel sounder}
\acrodef{CIR}[CIR]{channel impulse response}
\acrodef{CTF}[CTF]{channel transfer function}
\acrodef{DPM}[DPM]{Dirichlet process mixture}
\acrodef{TOA}[TOA]{time of arrival}
\acrodef{GMM}[GMM]{Gaussian mixture model}
\acrodef{OOK}[OOK]{on-off keying}
\acrodef{MLE}[MLE]{maximum likelihood estimation}
\acrodef{LOS}[LOS]{line-of-sight}
\acrodef{NLOS}[NLOS]{non-line-of-sight}
\acrodef{SG}[SG]{signal generator}
\acrodef{SA}[SA]{spectrum analyzer}
\acrodef{FDSOI}[FDSOI]{fully depleted silicon on insulator}
\acrodef{OpEx}[OpEx]{operational expenditures}
\acrodef{TCO}[TCO]{total cost of ownership}
\acrodef{CapEx}[CapEx]{capital expenditures}
\acrodef{MAC}[MAC]{medium access control}
\acrodef{GEO}[GEO]{geostationary orbit}
\acrodef{SWaP}[SWaP]{size, weight, and power}
\acrodef{NOMA}[NOMA]{Non-orthogonal multiple access}

\ifCLASSINFOpdf
\else
\fi

\begin{document}
\title{A Holistic Investigation on Terahertz Propagation and Channel Modeling Toward Vertical Heterogeneous Networks
}

\author{K{\"{u}}r{\c{s}}at~Tekb{\i}y{\i}k,~\IEEEmembership{Student Member,~IEEE,} Ali~R{\i}za~Ekti,~\IEEEmembership{Member,~IEEE,} G{\"{u}}ne{\c{s}}~Karabulut~Kurt,~\IEEEmembership{Senior~Member,~IEEE,} Ali~G\"{o}r\c{c}in,~\IEEEmembership{Senior~Member,~IEEE,} Halim~Yanikomeroglu,~\IEEEmembership{Fellow,~IEEE}

\thanks{K. Tekb{\i}y{\i}k and G.K. Kurt are with the Department of Electronics and Communications Engineering, {\.{I}}stanbul Technical University, {\.{I}}stanbul, Turkey, e-mails: \{tekbiyik, gkurt\}@itu.edu.tr}

\thanks{K. Tekb{\i}y{\i}k, A.R. Ekti, and A. G\"{o}r\c{c}in are with Informatics and Information Security Research Center (B{\.{I}}LGEM), T{\"{U}}B{\.{I}}TAK, Kocaeli, Turkey, e-mails: \{kursat.tekbiyik, aliriza.ekti, ali.gorcin\}@tubitak.gov.tr}

\thanks{A.R. Ekti is with the Department of Electrical--Electronics Engineering, Bal{\i}kesir University, Bal{\i}kesir, Turkey, e-mail: arekti@balikesir.edu.tr}

\thanks{A. G\"{o}r\c{c}in is with the Department of Electronics and Communications Engineering, Y{{\i}}ld{{\i}}z Technical University, {\.{I}}stanbul, Turkey, e-mail: agorcin@yildiz.edu.tr}
 
\thanks{H. Yanikomeroglu is with the Department of Systems and Computer Engineering, Carleton University, Ottawa, Canada, e-mail: halim@sce.carleton.ca}
 
}

\IEEEoverridecommandlockouts

\maketitle

\begin{abstract}
User-centric and low latency communications can be enabled not only by small cells but also through ubiquitous connectivity. Recently, the \ac{V-HetNet} architecture is proposed to backhaul/fronthaul a large number of small cells. Like an orchestra, the \ac{V-HetNet} is a polyphony of different communication ensembles, including \ac{GEO}, and \ac{LEO} satellites (e.g., CubeSats), and \acp{NFP} along with terrestrial communication links. In this study, we propose the \ac{THz} communications to enable the elements of \acp{V-HetNet} to function in harmony. As \ac{THz} links offer a large bandwidth, leading to ultra-high data rates, it is suitable for backhauling and fronthauling small cells. Furthermore, \ac{THz} communications can support numerous applications from inter-satellite links to in-vivo nanonetworks. However, to savor this harmony, we need accurate channel models. In this paper, the insights obtained through our measurement campaigns are highlighted, to reveal the true potential of THz communications in \acp{V-HetNet}. 
\end{abstract}

\IEEEpeerreviewmaketitle
\acresetall

\section{Introduction}\label{sec:intro}

Flexible wireless communication architectures including small cells, cell-free designs, \acp{hetnet}, and multi-band connectivity are widely aspired by the telecommunication industry for the user-centric system designs toward 6G and beyond to provide the required \ac{QoS} levels among users~\cite{yang_6g_2019}. Even though \ac{FSO} and mmWave technology have been discussed over the years, they are not capable of making this dream possible on their own. \ac{THz} wireless communication will be a critical enabler for 6G since it proposes a solution for the unlicensed band scarcity below $100$ GHz and it can be employed by not only macro-scale networks (e.g., inter-satellite links) but also by nano-scale networks (e.g., nanonetworks)~\cite{tekbiyik2019terahertz}. The small cell concept has not been fully deployed yet because of the backhaul and fronthaul cost. It is thought that along with the development of non-terrestrial networks, the pervasive connectivity can be provided, as well as the backhaul and fronthaul required by small cell and cell-free topologies~\cite{alzenad_fso-based_2018}. This 3D-network structure, depicted in \FGR{fig:vnets}, is termed as the \ac{V-HetNet}. As \ac{THz} communications can offer an enormous bandwidth enabling ultra-high data rates, it stands out among the competing solutions that are likely to be used in \acp{V-HetNet}. Although the transceiver costs for \ac{THz} communication are currently high, a cost reduction is expected with the developments in semiconductor technologies. 

This study presents an overview of the \ac{THz}-enabled \acp{V-HetNet} framework. Below, we explain the motivation behind the \ac{THz} communications in \ac{V-HetNet} and address challenges. By interpreting \ac{THz} channel behavior with measurement results, tips are given on how the system design and requirements will be affected by channel behavior in \acp{V-HetNet}. Finally, we pinpoint the open issues for the realization of \ac{THz}-enabled \acp{V-HetNet} by considering the measurement results.

\begin{figure*}[!t]
    \centering
    \includegraphics[width=\linewidth, page=2]{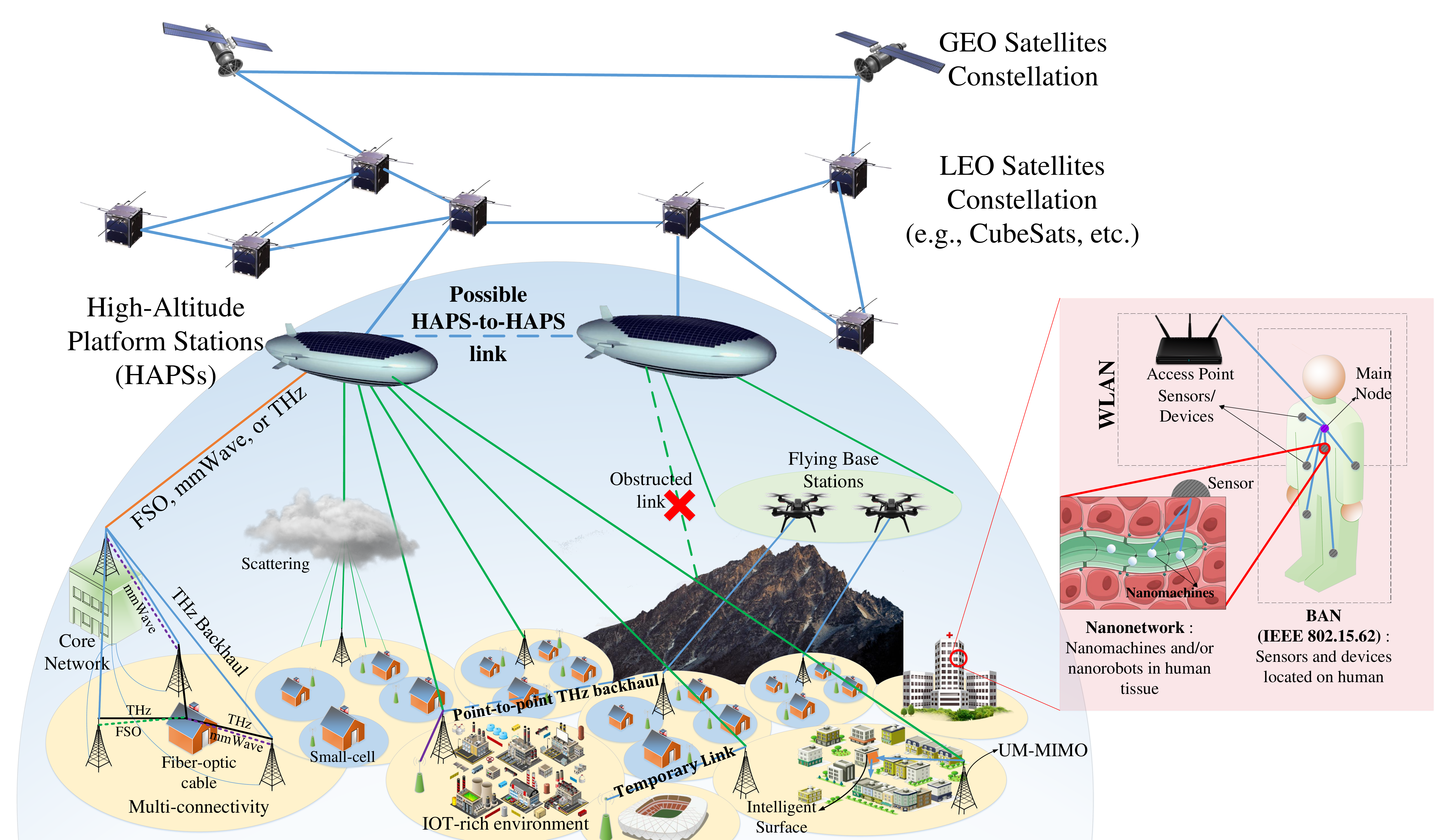}
    \caption{The \ac{THz}-enabled \ac{V-HetNet} framework including \ac{GEO}, and \ac{LEO} satellites, and \acp{NFP} along with terrestrial communication links. Furthermore, \ac{THz} networks will be a crucial enabler for the nanonetworks due to antenna size, while enabling the operation of \acp{hetnet} in harmony to support e-health applications.}
    \label{fig:vnets}
\end{figure*}

\section{Why Terahertz band?}\label{sec:thz_motivation}

Besides being a candidate technology for 6G, \ac{THz} communications imply a remarkable potential for the future of wireless communications as detailed below.

\subsection{Utilizing the not Fully-explored Frequencies}
 There is a need to push the carrier frequency beyond the mmWave band to overcome the spectrum scarcity below $100$ GHz. The \ac{THz} band allows wireless communication through tens of GHz bandwidth. Even though the path loss increases with frequency, the antenna gain is related to the square of the operating frequency. Therefore, high gain antennas can cope with the extreme path loss in the \ac{THz} band. As known, antennas with high operating frequency have relatively narrow beams compared to lower frequencies; hence, \ac{THz} antennas can be utilized in point-to-point links to backhaul small-cells. Moreover, since the antenna size decreases with frequency, it is possible to place a massive number of antennas on small surfaces. For example, it is possible to cover $1$ $\mathrm{mm}^{2}$ area with four antennas at $300$ GHz. Beyond the \ac{mMIMO}, namely \ac{UM-MIMO}~\cite{akyildiz_combating_2018}, can be employed to leverage the communication distance as well as the data rate enhancement. Although today's technology is in scarcity for high power \ac{THz} transmitters, it is expected that signal generators with high power can be created in the near future considering studies on graphene and InGaAs mHEMT MMICs. 

\subsection{Revolutionary Network Design Towards 6G}

\subsubsection{Cell-free Networking}
It is widely acknowledged that an architectural revolution is needed in 6G and beyond~\cite{chen_vision_2020}. Tiny cells and a cell-free architecture are demanded to cater the data-hungry end-users. In this regard, the high path loss, which is the main drawback of the \ac{THz} band, can be turned to account for the cell-free design. In other words, the \ac{THz} band is the region with the highest frequency reuse factor. Thus, it can be concluded that the \ac{THz} band is ideally suited for the cell-free networking. Not only backhaul from \acp{AP} to core network but also links between \ac{AP} and user equipments can utilize \ac{THz} waves to build up a cell-free network and satisfy the high data rate for backhaul. To support cell-free networking, a user-centric approach can be adopted for the spatial multiplexing by using \ac{UM-MIMO}. Additionally, distributed \acp{AP} are required fronthauling to the central processing unit. Therefore, cell-free networking can utilize the \ac{THz} waves to avoid fiber construction to every \ac{AP}.

\subsubsection{Vertical Heterogeneous Networks}

Recently, since aerial \acp{AP} such as \acp{UAV}, \acp{NFP}, and \acp{HAP} attract considerable attention, there has been an unprecedented vertical directional expansion in the wireless network. Furthermore, it seems that data flow between satellites and earth will increase more than ever in the coming years~\cite{akyildiz_internet_2019}. Although a dense satellite network has started to be established with the Starlink project and the Keppler project, the studies in this area are still nascent. \ac{FSO} and mmWave technologies have been proposed for the inter-satellite and satellite-to-X links. However, mmWave can only serve through a total of $9$ GHz bandwidth of the available unlicensed bands. On the other hand, as CubeSats are designed for \ac{LEO} operation, they need to move with a velocity of $28\times10^{3}$ kph to keep their orbits. This velocity necessitates fast beam steering; hence, \ac{FSO} with a mechanical gimbaled beam steering mechanism is susceptible to track the moving target. However, one face of $1$U CubeSat can be theoretically covered by $10^{4}$ antennas operating at $300$ GHz. Furthermore, to the extent allowed by the flight dynamics, the bottom of the \ac{NFP}'s wings can be covered with hundreds of antenna. Therefore, \ac{THz} enables to fast track a moving receiver with the help of electronic steering.

\ac{THz} \ac{mMIMO} setups can be grouped to serve a different number of users according to demand as illustrated in \FGR{fig:mmimo}. These enable a flexible communication network that can be dynamically shaped in conformity with the need~\cite{lin2016terahertz}. For example, the capacity demand possibly exceeds the target demand when event times or stadiums serve as an emergency assembly area. Here, \ac{THz}-aided \acp{HAP} can be used as a flexible communication structure to meet the increasing demand. Given the effort of each new generation to make the network more flexible, \ac{THz} communications has the potential to be a crucial enabler for 6G and beyond. 
\begin{figure}[!t]
    \centering
    \includegraphics[width=0.9\linewidth, page=3]{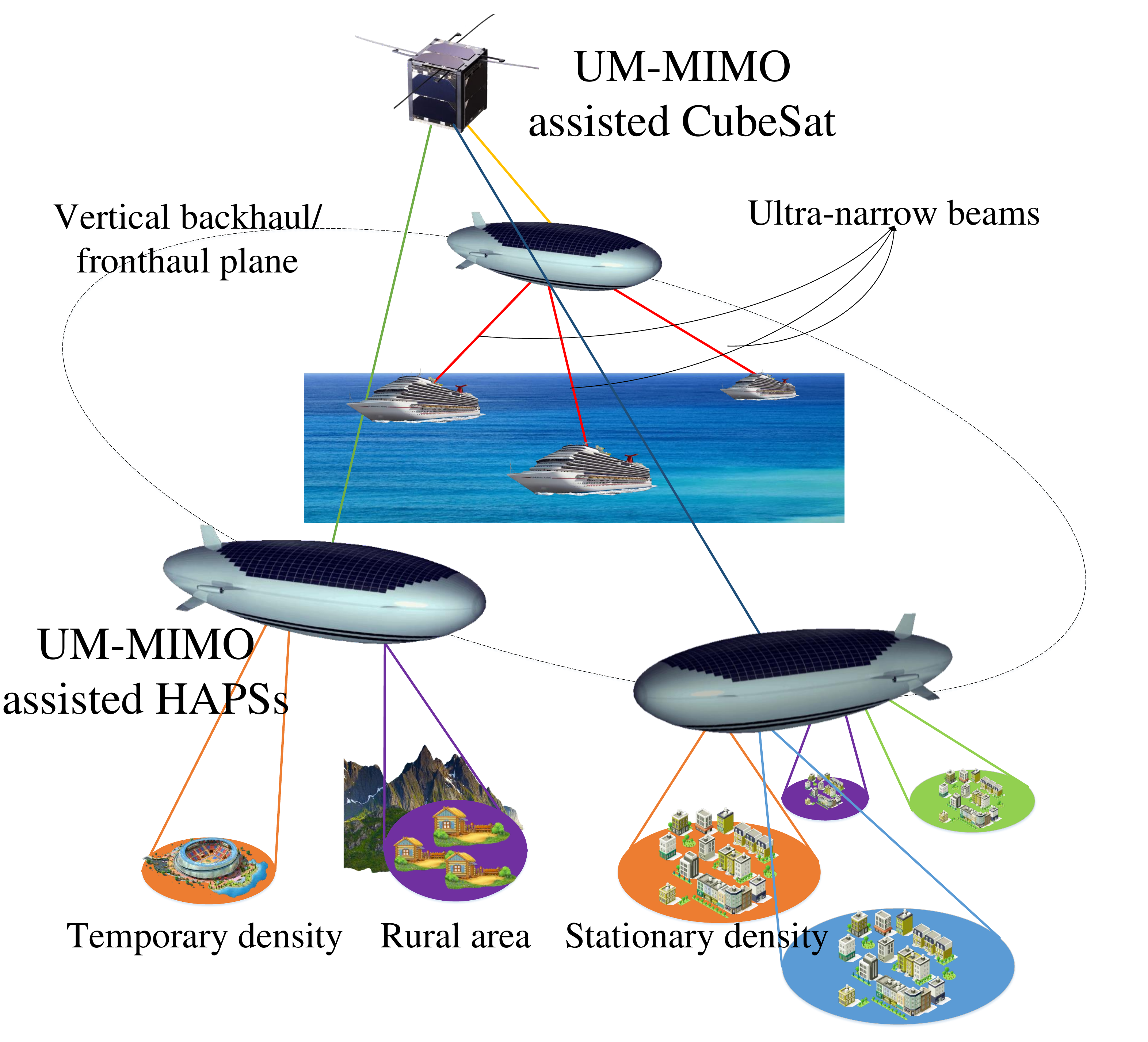}
    \caption{\ac{THz} networks can employ ultra-massive antenna arrays; hence, it can be turned into an advantage by dynamically adjustable beam pattern and multiple spot flexibility.}
    \label{fig:mmimo}
\end{figure}

\subsubsection{Multi-connectivity}
Ultra-reliable and low-latency communications in 5G require a high \ac{QoS} due to stringent requirements on availability and latency. Diversity and network availability increase as the user communicates with more than one \ac{AP} to improve \ac{QoS}. We infer that 6G and beyond systems can employ different technologies to enhance the \ac{QoS} levels. For example, in the \ac{THz} band, different sub-bands/windows can be used for communication at the same time, or complementary technologies such as mmWave, \ac{FSO}, and \ac{THz} can be used simultaneously.

\subsection{Applications from Planetary-scale to Nano-scale}

In the \ac{V-HetNet} framework toward 6G and beyond, it is envisioned that the world, which was previously connected by ground and undersea fiberoptic cables, is now connected over a wide space network. The near future communication technology pushes the limits of science-fiction. With the development of e-health technologies, billions of nanodevices and sensors need to be connected to the medical center over a unified and seamless network to enable low-latency communications. It is envisaged that using the \ac{THz} band in \acp{V-HetNet} will be useful to meet the latency and data rate required by e-health and haptic applications due to extremely short \ac{THz} pulses on the order of femtoseconds~\cite{jornet_femtosecond-long_2014}. For example, the \ac{THz} band paves the way for unprecedented applications initially suffering from the \ac{SWaP} constraints. The size of nanomachines can only be satisfied by the \ac{THz} antennas. As illustrated in \FGR{fig:vnets}, \ac{THz} waves can be employed in the harmony of nanonetworks, \ac{BAN}, and \ac{WLAN}. For instance, nanomachines communicate with each other and sensors with \ac{THz} waves. \ac{BAN} can utilize either IEEE 802.15.62 or mmWave (e.g., WiGig). The backhaul of \ac{AP} can be a directional \ac{THz} beam. 

\vspace{-0.2cm}
\subsection{Economic Feasibility}

To achieve resilient and reliable connectivity, mmWave and \ac{THz} band connection are needed as described in \SEC{sec:thz_motivation}. Furthermore, the mmWave technology market in the telecommunication industry, which is anticipated to reach $\$10.92$ billion by 2026 with a growing compound annual growth rate of $36.3\%$ per year, along with the corresponding industrial ecosystem, makes it attractive for the network operators and semiconductor companies \cite{terahertz_market}. Therefore, adopting mmWave and \ac{THz} band communication will stimulate the economy along with supporting Tbps data rates, novel usage scenarios, and applications. Thus, network operators exhibit particular interest in the proper adoption of the \ac{THz} band communication, which can further reduce the total cost of ownership, capital expenditures, and operational expenditures. Deploying a macro base station in a crowded and populated location such as downtown and providing seamless high-speed data rate connectivity to the remote islands and rural/suburban areas can be quite challenging and also very expensive due to the harsh geographical conditions. Thus, extending the wireless connectivity with the help of existing infrastructure to the areas beyond the reach of the fiberline by using the \ac{THz} band is anticipated to reduce the total cost. First, no additional or significant upgrades are needed for the backhaul traffic by the network operators since already existing infrastructure will be used for the traffic offloading which will ease the traffic congestion problem. Second, as for the software and hardware of the radio access network, there is no need for extreme modifications. Furthermore, the recent advances in the semiconductor technology enable to design at \ac{THz} frequencies; thus, antenna and chip can be integrated into the same package to reduce the production cost.

\section{Operational Challenges}\label{sec:challenges}

Attractive properties of \ac{THz} wireless communication can make it possible to build up the wireless communication network of the future. Nevertheless, there are many challenges to overcome. The state-of-the-art semiconductor technology cannot generate high power \ac{THz} waves. However, some promising transceiver designs have been proposed. It seems that a high output power transceiver will be designed by discovering the essence of graphene and \ac{CMOS}. Although the transmit power of \ac{CMOS} power amplifiers is about $1$-$10$ mW, they are in line with \ac{SWaP} constraints~\cite{tang2016cmos}. Fortunately, it is possible to design high gain antennas in this spectra to recover the signal power. 

In \acp{V-HetNet}, we do not expect \ac{THz} waves to suffer a severe loss in inter-satellite communication. The main problem is the operating temperature of the satellites. The temperature fluctuates concerning the position of Earth and Sun. Satellites already have a \ac{TCS} to maintain stable operation against changes in operating temperature. However, it may require to review the design of \ac{TCS}. Recently, NASA-JPL has pushed receiver design limits toward CubeSats. \ac{THz}-enabled satellite communication in the near future is not pie in the sky.

Since \acp{HAP} fly above the troposphere, which $99\%$ of the water vapor in the atmosphere is at this layer, no severe path loss is expected between \acp{HAP} and \ac{LEO} satellites. But, the relatively high speed between \acp{HAP} and satellites, and the sharp beam of \ac{THz} antennas require stringent tracking of transmitter and receiver. The electronic steering functionality of \ac{UM-MIMO} can provide fast and accurate tracking with the cost of large antenna arrays.

A high attenuation rate is observed between \acp{HAP} and ground stations since water vapor absorbs most of the power of the \ac{THz} wave, and the region where the water vapor is most dense in the atmosphere is up to $20$ km above the Earth. The atmospheric attenuation coefficient at $300$ GHz is around $3$ dB/km and $30$ dB/km for very light and dense humidity levels, respectively. The attenuation coefficients imply that the \ac{THz} wave is more advantageous than \ac{FSO} in terms of path loss~\cite{alzenad_fso-based_2018, siles_atmospheric_2015}.

Another challenge lies in \ac{UAV} communication. \acp{UAV} with high mobility vibrate slightly, even when they are suspended in the air, due to flight aerodynamics. This poses a serious challenge to \ac{THz} communication, which has a very narrow antenna beam. The loss due to misalignment between \ac{THz} antennas is known to be severe~\cite{ekti2017statistical}. The beam search may take a long time after the antenna alignment deteriorates. At this point, we think that it would be appropriate to make \ac{THz} communication with wider beams using beam broadening. Also, the phased array can be adaptively controlled with the help of a Kalman filter and gyroscope on the \ac{UAV}; however, this can cause energy scarcity for battery-dependent \acp{UAV}. 

As high data rate \ac{THz} pulses have very short-times; multipath components may not overlap with the first-path component. However, the propellers' periodic behavior may give rise to rotor modulation. As the wavelength is much smaller than propellers, a simple blockage model can be utilized to analyze the effect of the rotor modulation. 

There are additional challenges that may occur in each layer of \ac{V-HetNet}. These include beam steering, high-performance computation, and low-complexity tracking. Also, ultra-large antenna array configuration, mutual coupling among antennas, and correlation among sub-channels should be jointly considered in the design of \ac{UM-MIMO}~\cite{tekbiyik2019terahertz}. One of the main challenges is to estimate the propagation channel for \ac{UM-MIMO} \ac{THz} communication. Since it is expected to use \ac{UM-MIMO} systems with thousands of antennas, novel intelligently designed channel acquisition algorithms are needed. \ac{DL}-based channel estimation techniques can be utilized ubiquitously. The sparse nature of \ac{THz} wireless channels allows using compressed sensing so that the complexity of acquisition methods can be reduced.

\section{Overview of Measurement Studies}

\subsection{Channel Modeling Methodologies}\label{sec:thz_ch_modeling} 

Wireless channel modeling is mainly categorized under two approaches: deterministic and stochastic methods. The deterministic methods are propagation medium-specific. They require the model of the environment in-depth with the material characteristics and dimensions. Although an accurate channel model is obtained for a given environment, the computation complexity is extremely high, and it exponentially increases with the dimensions of the environment. Ray-tracing is the most used deterministic technique for channel modeling. Ray-tracing models the scattering, diffraction, and reflection effects in with geometric optics. Finite-difference time-domain, \ac{MoM}, and \ac{FEM} are some of the other deterministic channel modeling techniques. Since \ac{MoM} and \ac{FEM} are working in the frequency domain, these methods do not seem to be the best approach as they require excessively complex computations in channel modeling for a huge \ac{THz} band, but geometric optics that become more accurate due to the stronger corpuscular behavior in the \ac{THz} band.

Unlike deterministic modeling, stochastic channel modeling aims to create a channel model independent of the environment as much as possible. For this purpose, a statistical average of many measurements is used to create the channel model. For the stochastic channel modeling, measurements are usually carried out with a \ac{VNA} or \acl{CS}. The main difference between these devices is the domain where they perform operations. Channel sounder computes the channel impulse response by using an uncorrelated m-sequence, but \ac{VNA} composes many narrowband measurements to create \acl{CTF}. \ac{VNA} performs more accurate channel modeling due to calibration for each narrowband and encountering low noise, but the measurement time is long~\cite{priebe2013ultra_black}. Another difference is that \ac{VNA}-assisted measurement does not require extra clock synchronization due to two ports in the same device, whereas \acl{CS} requires a strict external clock to sync two distinct ports.

Due to the presence of the colossal bandwidth, multipath components can be distinguished. The nature of the \ac{THz} band leads to severe frequency selectivity. As a result, conventional small scale flat fading models are not appropriate, and the broadening effect cannot be ignored.

\subsection{Measurement Setup and Results}\label{sec:measurement}

In this section, we detail the \ac{THz} measurement campaigns and results. Firstly, \ac{VNA}-assisted measurements are carried out to observe \ac{THz} channel behavior with high spectral resolution. Afterward, pulse-based measurements for short-range \ac{THz} communication will be addressed since \ac{OOK} is seriously envisaged for use in \ac{THz} wireless communications~\cite{jornet_femtosecond-long_2014}. The \ac{VNA}-assisted measurement provides high spectral resolution at the cost of long measurement time while the pulse-based method lacks high resolution, but it allows taking measurements in a short time.

\subsubsection{VNA-assisted Channel Modeling}
An experimental measurement setup is constructed in the anechoic chamber with dimensions of $7\mathrm{m}\times3\mathrm{m}\times4\mathrm{m}$ as seen in \FGR{fig:measurement_setup}. To eliminate the specular effects, the perimeter of the system is covered with absorbent material. The measurement setup consists of a \ac{VNA}, WR-03 waveguided extension modules, and an extender controller. WR-03 extension modules expand the \ac{VNA}, whose upper limit is $67$ GHz, to \ac{THz} band by multiplying RF input signal with $18$. In cases where an extender module is used, it should be preferred to use as few frequency multipliers as possible. This is observed due to the increasing number of striking, phase distortion, and intermodulation product effect increases. A single multiplier, if possible, is preferred to avoid mentioned distortions. The extender module employed in this setup includes single frequency multiplier.
\begin{figure}[!t]
    \centering
    \includegraphics[width=\linewidth]{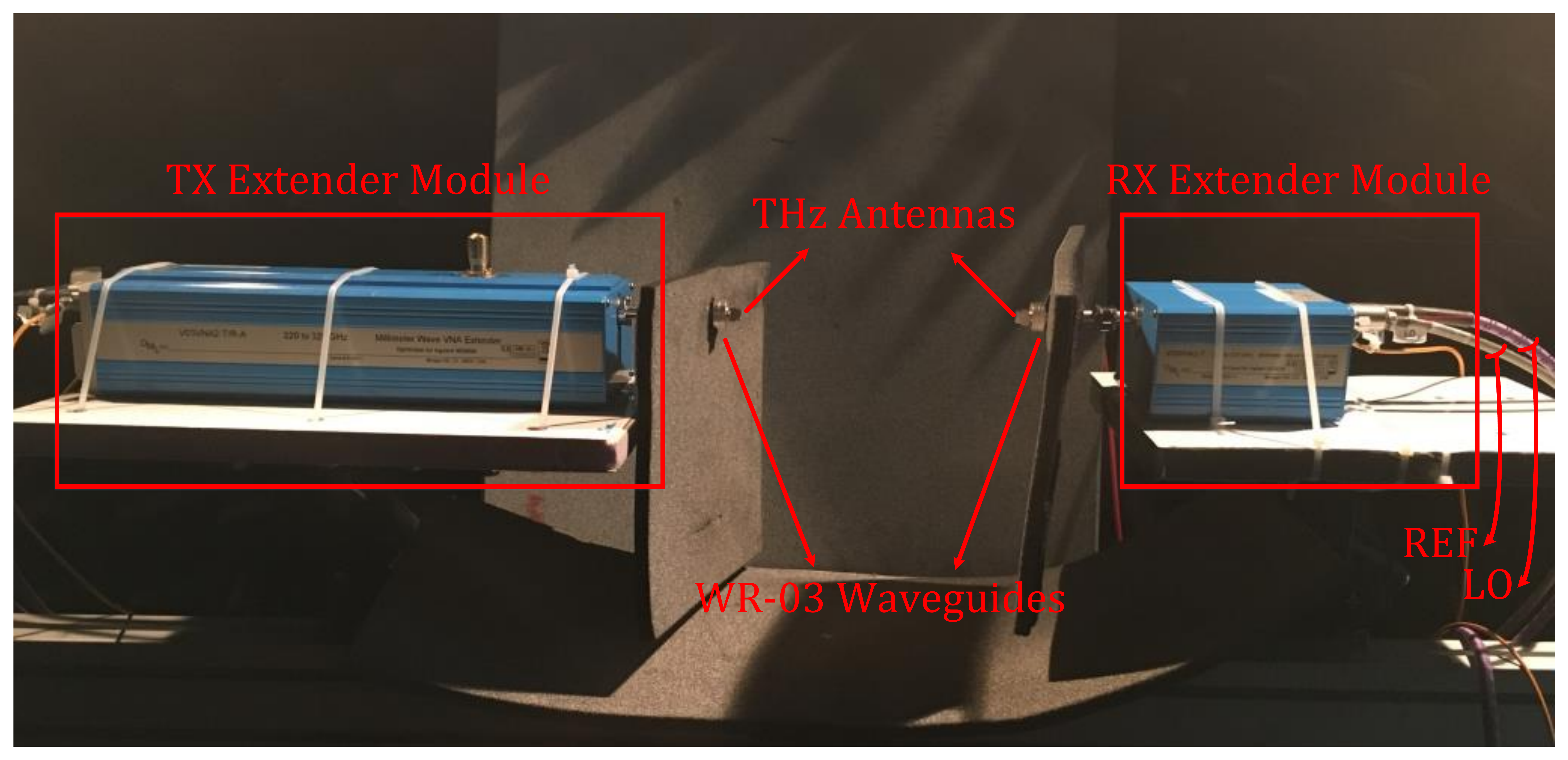}
    \caption{The measurement setup consists of \ac{VNA} equipped with \ac{THz} extender modules with WR-03 waveguide, and extender controller.}
    \label{fig:measurement_setup}
\end{figure}

Driving \ac{VNA} with the input signal in between $13.34$ GHz and $16.67$ GHz results in \ac{THz} waves with carrier frequencies in between $240$ GHz and $300$ GHz. To protect the phase and amplitude stability, instead of using all the band provided by the extender, measurement is taken in the band between $240$ and $300$ GHz by not using the edge points of the band~\cite{ekti2017statistical}. The IF signals downconverted at transmitter and receiver are used to find \acl{CTF} by comparing them to acquire scattering parameters (i.e., $S_{21}$). \ac{VNA}-assisted channel modeling provides high-frequency resolution by taking measurements. $4096$ frequency points are employed within the operation band demarked by the \ac{VNA}. As a result, $14.648$ MHz spectral resolution is provided by the measurement setup. 

By using measurement results, the frequency dependency of the path loss is revealed in \cite{ekti2017statistical}. The measurement results indicate similar channel behavior at varying distances. However, the loss in some frequency intervals is higher than in others. For example, there is a noticeable reduction in the power of the received signal between $270$ GHz and $290$ GHz. As stated above, antenna alignment is crucial for \ac{THz} communication because of the relatively narrow beamwidth. It is shown that the pointing error in antenna alignment gives rise to a sharp fall in the received power. Since the received power decreases exponentially, it is reasonable to model this decay as an exponential distribution. The pointing errors between receiver and transmitter are pointed as challenges in \SEC{sec:challenges}. Thus, how pointing error degrades the received power is investigated by tilting the antennas. As seen in \FGR{fig:cir_80cm_alignment}, the antenna misalignment causes a decrease in the received peak power. The power decrease concerning the peak power corresponding to the \ac{LOS} path is approximately $2.3$ dB and $13$ dB for $10^{\circ}$ and $20^{\circ}$ tilts, respectively. These results clearly demonstrate the need for precise alignment between the transceivers. Similar to~\cite{siles_atmospheric_2015}, the humidity has no significant effect on the received signal power. In the light of measurement results, it is proved that \ac{THz}-enabled \acp{V-HetNet} strictly need an accurate resolution in the beam steering as well as beam broadening before fine-tuning. However, considering that the eavesdroppers cannot be aligned with the transmit antenna due to the narrow beam, power loss due to misalignment increases communication security.

\begin{figure}[!t]
    \centering
    \includegraphics[width=\linewidth]{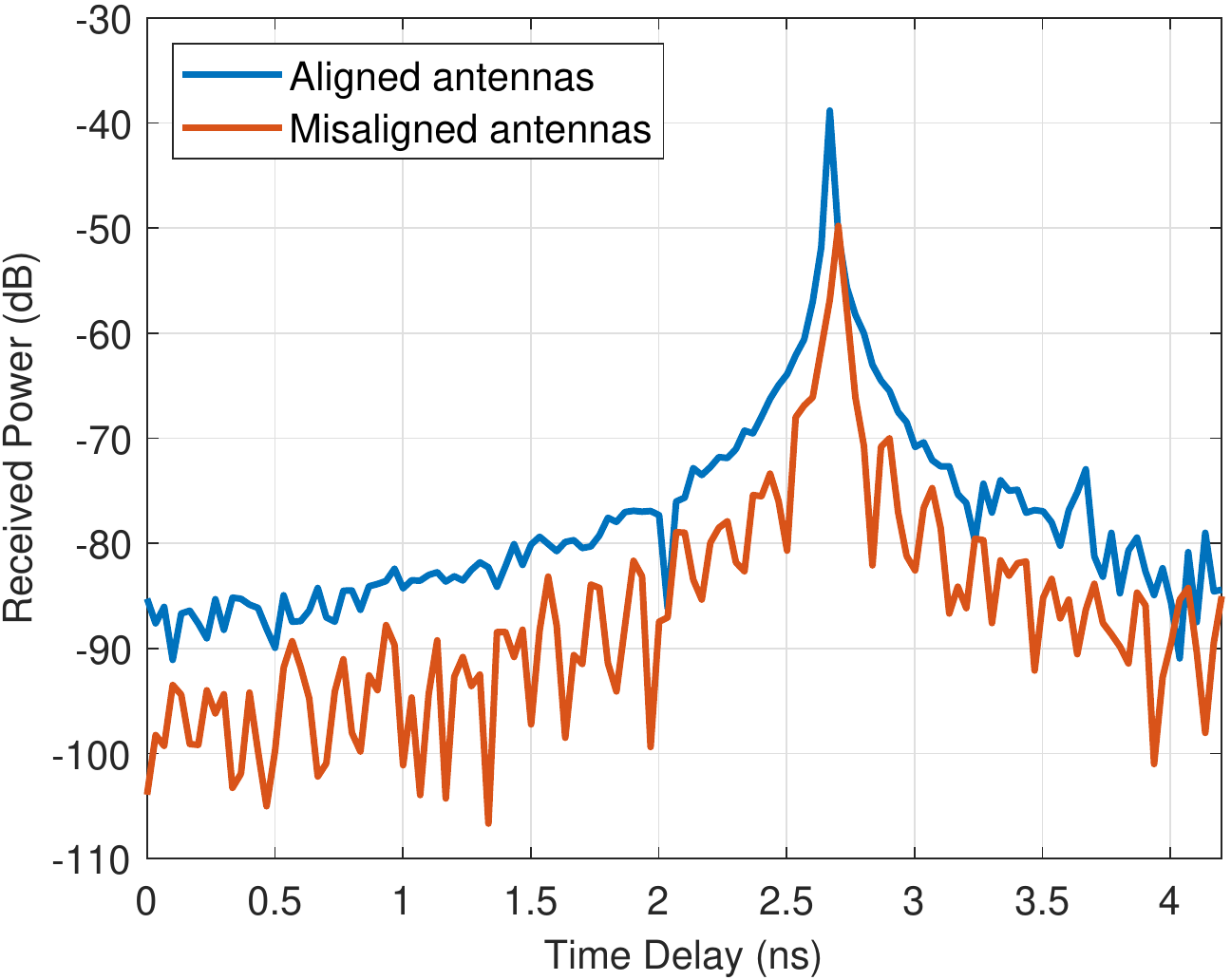}
    \caption{The \ac{VNA}-assisted measurements show the severe effect of misalignment on the received signal power which decreases $13$ dB the in case of $20^\circ$ tilt.}
    \label{fig:cir_80cm_alignment}
\end{figure}

\subsubsection{Pulse-based Channel Modeling}
As \ac{OOK} modulation is mainly considered for \ac{THz} communication, specifically in nanonetworks, we prepare a measurement campaign to analyze the channel behavior for a pulse train~\cite{tekbiyik2019statistical}. The measurement setup detailed above is used without a change except for using a \acl{SG} and a \acl{SA} rather than \ac{VNA}. As two ports are on distinct devices, the synchronization is ensured by connecting the reference clock of the \acl{SG} to the external clock input of the \acl{SA} via a cable. The band between $275$ GHz and $325$ GHz is measured by feeding the extender module with a pulse train generated by the \acl{SG}. The width of the pulse is limited with a minimum of $140$ ns due to signal generator capability. At the receiver side, the pulse train is recorder during $1$ms by taking $100,000$ I/Q samples. The study focuses on path loss relations with frequency and distance. In this regard, \FGR{fig:pulse_power} denotes both frequency and distance related path loss. The same pattern through the band is observed for each range.

Due to a frequency dependency, a linear path loss model may not be adequate in such a wide band. We denote that the path loss obeys the two-slope model~\cite{tekbiyik2019statistical}. It is the evidence that for \ac{THz} the distance is another determinant along with the frequency. By employing the two-slope model, the normalized received power is illustrated in \FGR{fig:log_avg_normalized_rx_power} for the unit power transmitter at $300$ GHz. These results imply that the distance-dependent channel characteristics require distance-aware \ac{MAC} to control physical layer parameters such as modulation order and coding rate.

The high frequency reuse factor due to high path loss enables the design of small cell networks and cell-free architectures. The measurement results point that the most challenging part of \ac{THz}-enabled \acp{V-HetNet} is the link between \acp{HAP} and terrestrial \acp{AP} as the main path loss arises through the long distance, where the majority of water molecules in the atmosphere are found and the antenna alignment is compelling over the long-distance. Because of the frequency selectivity, the \ac{THz} band can be divided into sub-bands; thus, using different sub-bands can support the multiple connections in the \acp{V-HetNet}.

\begin{figure}[!t]
    \centering
    \includegraphics[width=\linewidth]{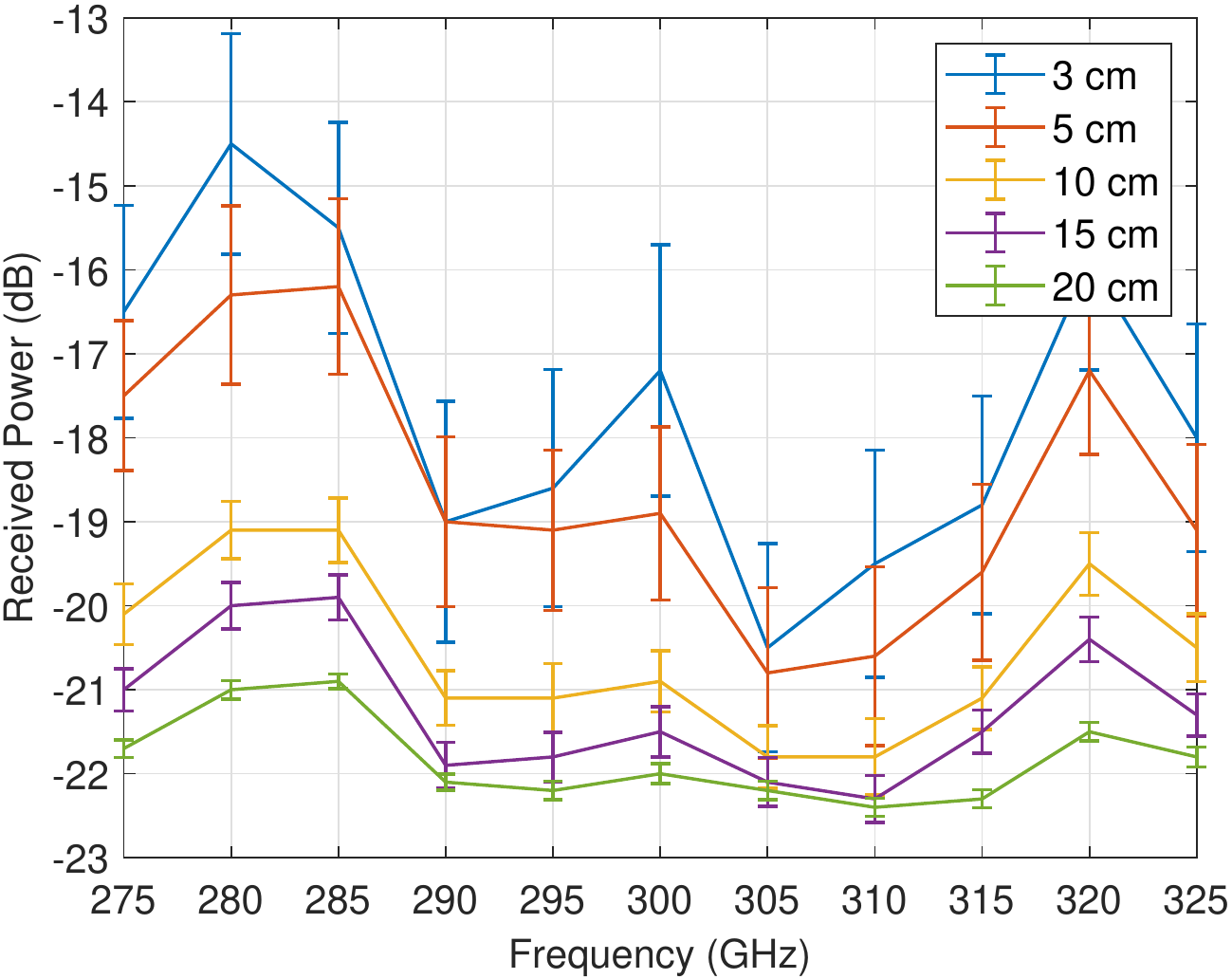}
    \caption{The received power with respect to frequency obtained by pulse train driven measurements. The $95\%$ confidence intervals are denoted with error bars.}
    \label{fig:pulse_power}
\end{figure}

\section{Open Issues and Research Directions}\label{sec:open_issues}

Owing to the fact that \ac{THz} is one of the key enablers of \acp{V-HetNet}, in this early stage of the technology, there are many open issues to be addressed to make it viable.

\paragraph{Cell-free Networking} To construct the user-centric network of the future, it is required to satisfy a decentralized seamless connection. Therefore, promising tools such as radio stripes, metasurfaces (e.g., reconfigurable intelligent surfaces), and holographic beamforming are needed to meet \ac{THz} wireless communication. Moreover, \ac{DL} can be employed for the efficient operation of cell-free networks (e.g., controlling transmitter power in \ac{UM-MIMO}).

\paragraph{Software-defined and Self Organizing Networks} The software-defined network (SDN) and self-organizing network (SON) pave the way for not only ubiquitous connection but also low-cost network operation. SDNs and SONs require dynamic \ac{MAC}, spectrum sharing, and spectrum switching, \ac{DL}, and \ac{DRL} enable to improve the elasticity of networks by considering feedback from the medium to find optimal network policy owing to the fact that hundreds of billions of devices will be connected. \ac{DRL} and blockchain are promising techniques to meet this demand~\cite{dai2019blockchain} by utilizing intelligent spectrum sharing protocols. Moreover, federated learning allows to distribute the computation cost over the thousands of the connected devices, and therefore the central processing unit may not be needed. Multi-band antennas and RF chains are also required to support spectrum switching. 

\begin{figure}[!t]
    \centering
    \includegraphics[width=\linewidth]{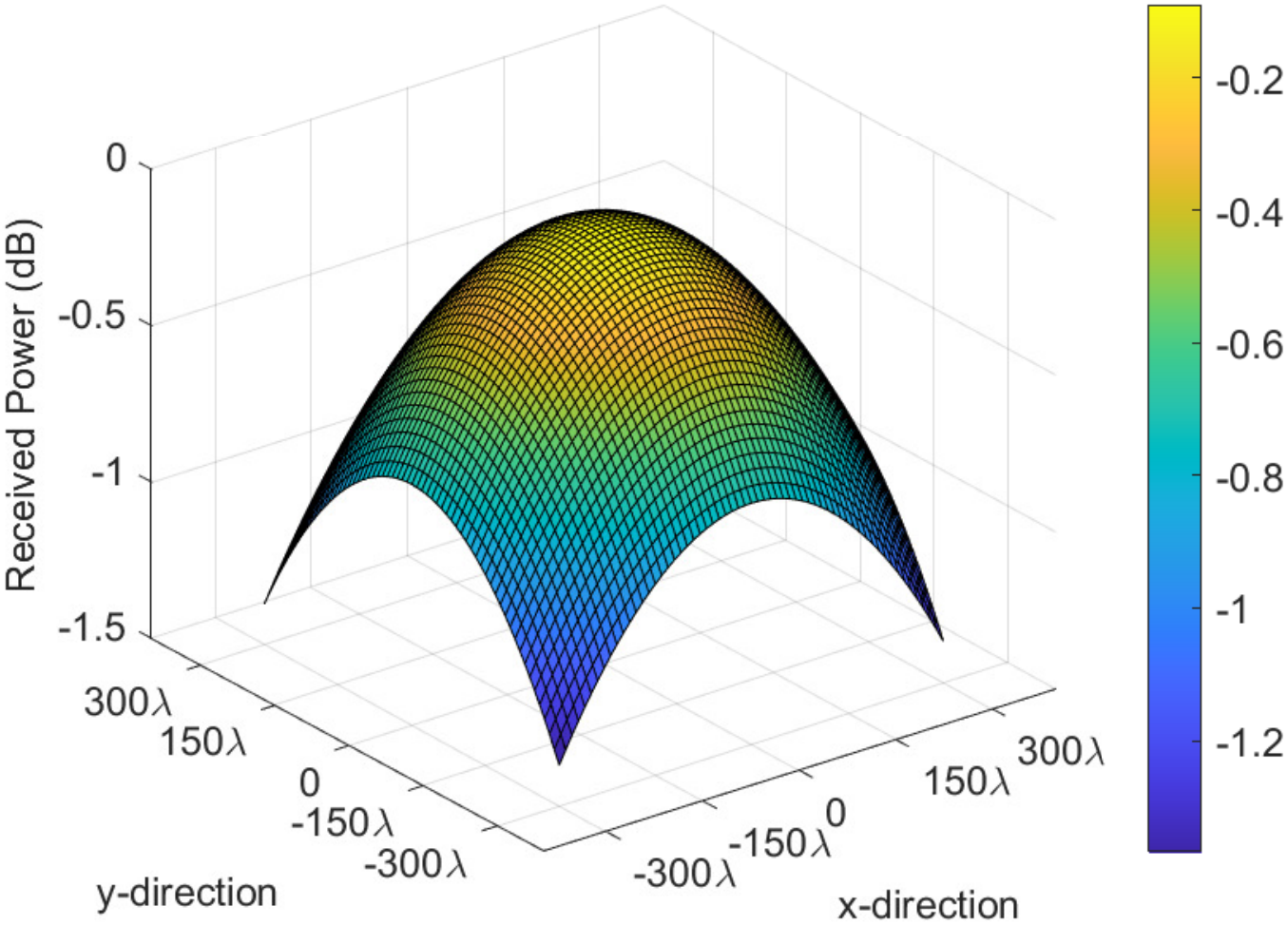}
    \caption{The normalized received power considering the two-slope path loss model with unit power transmitter.}
    \label{fig:log_avg_normalized_rx_power}
\end{figure}

\paragraph{Channel Modeling and Estimation} In the absence of the \ac{LOS} link, the communication may be carried out through a non-line-of-sight link; hence, reflection and scattering of \ac{THz} waves must be investigated in the \ac{V-HetNet} framework. Few studies model the spatial correlation in \ac{UM-MIMO} channels. Low-complexity methods with compressed sensing by utilizing the sparsity of \ac{THz} channels can be investigated. Also, it is expected that \ac{DL}-based approaches will be useful for understanding channel characteristics. For instance, generative adversarial networks are able to generate infinitely many channel states, so we can model the channel estimation methodology, which is due to the effect of fast changes in channel characteristics. It is noted that the weather conditions such as rain, fog, and clouds have crucial impacts on the channel characteristics. Therefore, they should be considered in the models. Although the channel modeling document by IEEE 802.15.3d Task Group points to some weather impacts on the channel, it needs further investigation. Furthermore, for the feasibility of the operations of \acp{HAP}, thorough airborne channel modelling activities are needed.

\paragraph{Multiple Access Techniques} The severe frequency selectivity plays an important role when choosing multiple access technology. For example, code-division multiple access suffers from high-cost equalizer to recover the received signal. However, we believe that subcarrier-based access technologies, like orthogonal frequency division multiple access, are suitable to employ in the \ac{THz} networks. Yet, it is seen that peak-to-average power ratio reduction methods are required owing to an excessive number of nodes. \ac{NOMA}, proposed to increase spectral efficiency, is likely to be used in \ac{THz} systems, but there is a trade-off between the additional complexity that \ac{NOMA} would bring and the increase in efficiency. Due to \ac{mMIMO} and pencil-sharp beams, the beam division multiple access is highly welcome in \ac{THz} communications. As in IEEE 802.11ad, single-carrier frequency division multiple access can be used for \ac{THz} communications.

\paragraph{Beam Alignment} It seems that utilizing \ac{UM-MIMO} in user equipment is formidable owing to the high computational complexity of beamforming algorithms and space limitations for antennas. Hence, a situation-aware \ac{MAC} layer is required, which allows adaptively utilization of \ac{THz} and mmWave. Further, intelligent surfaces can be key enablers for \ac{THz} user equipment because the computations are made on the medium rather than user equipment.

\section{Summary}\label{sec:conclusion}
 \ac{THz}-enabled \acp{V-HetNet} provide not only ubiquitous connection but also relatively low-cost dense networks. As projected, different wireless communication technologies, including \ac{FSO}, mmWave, and \ac{THz}, will be orchestrated for 6G and beyond; however, \ac{THz} technologies will be the maestro due to widespread application strategies. Howbeit, novel approaches considering channel behavior should be adopted from satellite to in-vivo \ac{THz} communications. 

\vspace{-0.2cm}

\bibliographystyle{IEEEtran}
\bibliography{ieee_commag_thz}

\vspace{-0.2cm}

\section*{Biographies}
\footnotesize{K{\"{U}}R{\c{Ş}}AT TEKBIYIK [StM'19] (tekbiyik@itu.edu.tr) is pursuing his Ph.D. degree in Telecommunication Engineering at Istanbul Technical University and is also researcher at TUBITAK BILGEM.\\

AL\.{I} RIZA EKT\.{I} received Ph.D. degree in Electrical Engineering from Department of Electrical Engineering and Computer Science at Texas A\&M University in 2015. He is currently an assistant professor at Balikesir University and also senior researcher at TUBITAK BILGEM.\\

G{\"{U}}NE{\c{Ş}} KARABULUT KURT [StM'00, M'06, SM'15] (gkurt@itu.edu.tr) received the Ph.D. degree in electrical engineering from the University of Ottawa, Ottawa, ON, Canada, in 2006. Since 2010, she has been with Istanbul Technical University. She is serving as an Associate Technical Editor of IEEE Communications Magazine.\\

AL\.{I} G\"{O}R\c{C}\.{I}N received his Ph.D. degree in University of South Florida. He is currently assistant professor at Yildiz Technical University in Istanbul and also manager at BILGEM BTE.\\

HALIM YANIKOMEROGLU [F] (halim@sce.carleton.ca) is a full professor in the Department of Systems and Computer Engineering at Carleton University, Ottawa, Canada. His research interests cover many aspects of 5G/5G+ wireless networks. His collaborative research with industry has resulted in 37 granted patents. He is a Fellow of the Engineering Institute of Canada and the Canadian Academy of Engineering, and he is a Distinguished Speaker for IEEE Communications Society and IEEE Vehicular Technology Society.
}

\balance
\end{document}